\documentclass[10pt,conference]{IEEEtran}

\usepackage{booktabs}
\usepackage{amsmath}
\usepackage{cite}
\usepackage{longtable}
\usepackage{multirow}
\usepackage{url}
\usepackage{xspace}
\usepackage{listings}
\usepackage{xcolor}
\usepackage[inline]{enumitem}
\usepackage{tikz}
\usetikzlibrary{arrows.meta, positioning, shapes.geometric, fit, backgrounds, calc}
\usetikzlibrary{arrows.meta}
\usepackage{tabularx}
\usepackage{caption}
\usepackage{balance}
\usepackage[hidelinks,implicit=false]{hyperref}
\usepackage{microtype}

\usepackage{inconsolata}

\newcommand{\tool}{BMC-Agent\xspace}
\newcommand{\fmagent}{FM-Agent\xspace}
\newcommand{\vibeos}{VibeOS\xspace}

\newcommand{\dynconf}{\textsc{confirmed\_dynamic}\xspace}
\newcommand{\sysconf}{\textsc{confirmed\_system\_entry}\xspace}
\newcommand{\bmcconf}{\textsc{confirmed\_bmc}\xspace}

\newcommand{\commentout}[1]{}
\newtheorem{lemma}{Lemma}
\newtheorem{definition}{Definition}
\usepackage[ruled,vlined,linesnumbered]{algorithm2e}
\newcommand{\lit}[1]{\textup{\textsc{#1}}}
\SetKwInput{KwIn}{Input}
\SetKwInput{KwOut}{Output}
\SetKwComment{tcc}{// }{}
\SetAlCapNameSty{textbf}
\SetAlFnt{\small}
\SetAlCapFnt{\small}
\SetAlCapSkip{0.4em}
\DontPrintSemicolon
\SetKwProg{Proc}{Procedure}{}{}
\SetKwFunction{Validate}{Validate}
\SetKwFunction{Refine}{Refine}
\SetKwFor{PForEach}{foreach}{in parallel do}{end}

\title{Agentic Model Checking}
\author{
\IEEEauthorblockN{
Youcheng Sun\IEEEauthorrefmark{1},
Jiawen Liu\IEEEauthorrefmark{1},
Humaid Alqasimi\IEEEauthorrefmark{1},\\
Lucas Cordeiro\IEEEauthorrefmark{2},
Daniel Kroening\IEEEauthorrefmark{3},
Chun Jason Xue\IEEEauthorrefmark{1}}
\IEEEauthorblockA{
\IEEEauthorrefmark{1}Mohamed bin Zayed University of Artificial Intelligence (MBZUAI)\\
\IEEEauthorrefmark{2}University of Manchester\quad
\IEEEauthorrefmark{3}Amazon\\
\href{mailto:Youcheng.Sun@mbzuai.ac.ae}{Youcheng.Sun@mbzuai.ac.ae},
\href{mailto:Jiawen.Liu@mbzuai.ac.ae}{Jiawen.Liu@mbzuai.ac.ae},
\href{mailto:Humaid.Alqasimi@mbzuai.ac.ae}{Humaid.Alqasimi@mbzuai.ac.ae}\\
\href{mailto:lucas.cordeiro@manchester.ac.uk}{lucas.cordeiro@manchester.ac.uk},
\href{mailto:dkr@amazon.com}{dkr@amazon.com},
\href{mailto:Jason.Xue@mbzuai.ac.ae}{Jason.Xue@mbzuai.ac.ae}}
}

\begin{document}

\maketitle

\begin{abstract}
Verifying large software systems is difficult.  Formal specifications are
rarely available, and the assumptions that code relies on are often encoded
implicitly at call sites rather than enforced at function boundaries.  This
paper presents \emph{agentic model checking}, a verification scheme that
combines LLM agents with bounded model checking (BMC) under the principle
that \emph{agents propose, solvers verify}: agents perform semantic
reasoning such as specification inference and verification planning, while
the formal backend checks the resulting bounded verification queries. We realize
this scheme in \tool, an agentic verification system built on an
off-the-shelf bounded model checker.

The key contribution of this paper is \emph{agentic orchestration} of the
model checker.  Rather than fixing a single verification recipe, \tool plans
how bounded model checking should be applied to each verification task.  It
infers function specifications, configures the verification task, decomposes
the analysis using specification-constrained stubs, validates
counterexamples through deterministic checks and replay, and iteratively
refines specifications in response to spurious witnesses.  This
orchestration enables the same BMC engine to scale from individual functions
to large software systems, preserving solver-backed verification.

We evaluate \tool on LLM-generated operating-system and compiler code,
standard verification benchmarks, and mature OSS-Fuzz-hardened libraries. 
Across these workloads, \tool discovers confirmed defects in heavily tested
software and produces bounded verification results without requiring manually
written specifications.  These results demonstrate that agentic
orchestration substantially extends the practical scalability of bounded
model checking for real-world software.
\end{abstract}


\section{Introduction}
\label{sec:intro}

LLM coding assistants~\cite{chen2021evaluating} now generate substantial bodies of systems
code, operating systems, compilers, drivers, and
embedded utilities, in both C and increasingly Rust.  Studies report a high
number of security-relevant defects~\cite{pearce2025asleep,
dai2025rethinking}.  Verifying the code is hard. The developers who deploy
it rarely understand every function, and manual specification is infeasible at
scale. The defects are often low level and 
require exhaustive reasoning over all inputs to rule out.  An LLM can
 flag a suspicious line; it cannot prove the bug absent.

A second pattern recurs across codebases. LLMs encode safety
contracts \emph{implicitly} at call sites rather than enforcing them at function
boundaries. Public byte-helpers (slice readers, offset-arithmetic writers) omit
bound and overflow guards, while every in-tree caller happens to maintain the
implicit invariant. The code works on well-formed input and fails on crafted
input through the same public functions, even though no current call site
triggers it. Hence, a verifier that reports errors without distinguishing in-tree
reachability from public-API ingress either over-reports or under-reports;
useful verification needs explicit framing of which threat model is applied and
which bugs are active versus latent.

The verification community has techniques for exactly these bugs.  Bounded
model checking (BMC)~\cite{biere1999symbolic} determines whether a property
is violated within an execution of at most $k$ steps; CBMC~\cite{cbmc}
implements it for C, Kani~\cite{kani-ase} for Rust, and JBMC~\cite{cordeiro2018jbmc}
for Java bytecode, each returning a concrete counterexample or a bounded
proof.  Compositional verification~\cite{assume-guarantee} scales analysis by
decomposing verification into component-level obligations. 
Counterexample-guided abstraction refinement (CEGAR)~\cite{cegar}
iteratively refines abstractions to eliminate spurious counterexamples. 
These techniques still need properties to check, and compositional
verification additionally relies on suitable component contracts.  Such
properties and contracts are often absent from LLM-generated code.  Recent
work addresses this gap using LLMs
themselves~\cite{fmagent, autospec, lemur, wu-invariants}.  LLM-only
approaches scale well and provide useful explanations, but are unsound for
arithmetic and memory-safety properties that require exhaustive path
reasoning.  LLM-assisted deductive verification provides stronger
guarantees, but typically produces no concrete counterexamples when
verification fails.

\paragraph*{Agentic model checking}
We combine LLM agents with a {bounded model checking} backend following
the principle \emph{agents propose, solvers verify}.  Agents do the semantic
work, inferring specifications, selecting which checks each function
warrants, classifying counterexamples, and proposing refinements.  These
outputs are then checked using specification parsing, BMC queries,
solver-based soundness guards, and concrete execution.

Our target is unannotated systems code. Classical model checking
usually starts with a property and a verification environment; here the
agents must construct them. That task faces two obstacles. Without
specifications, functions cannot be checked one at a time, and with many
functions the whole program cannot be encoded in one query. One artifact, the
per-function contract, addresses both obstacles.
Inference proposes the contract, configuration decides how much real code enters the
query, validation tests witnesses, refinement tightens it, and the result
re-enters verification as that callee's stub.  That loop, not any one stage,
is the contribution.

We choose BMC because its bounded queries check low-level memory and
arithmetic behaviour and yield concrete counterexamples for replay and
contract refinement. The intended outputs are validated defect witnesses
and verification results within explicit bounds, not unbounded proofs
of program correctness. This choice does not make the orchestration
principle specific to BMC, nor do we suggest that CEGAR-based verifiers cannot
produce concrete witnesses. Our specification-refinement loop draws on
CEGAR, but replacing the BMC backend with an abstraction-based verifier
would require a separate integration and evaluation, which we do not
claim here. Functions are checked separately against their specifications, using
postcondition-constrained stubs for abstracted callees. Independent
verification tasks can therefore run in parallel. Accepted contract
refinements are reused in subsequent verification of callers.
A strengthened callee precondition becomes an updated obligation
at each call site.

In summary, we
\begin{enumerate}[leftmargin=1.4em]

\item present \emph{agentic model checking} and implement it in \tool, where
LLM agents orchestrate an existing bounded model checker; 

\item evaluate \tool on standard verification benchmarks,
LLM-generated systems, and mature open-source software, comparing
it with traditional and LLM-based verification tools and reporting
confirmed defects and bounded verification results.

\end{enumerate}

\section{Motivating Example}
\label{sec:motivating}

LLMs can flag suspicious code. Verifying it often requires
recovering implicit contracts and checking whether callers establish
them. Listing~\ref{lst:packet} uses a byte-reading helper adapted
from the AI-generated project VibeOS~\cite{vibeos}. The surrounding
packet parser and call chain are constructed for this example.

\paragraph{Recovering the contract by refinement}
\texttt{read\_u32} reads four bytes without a bounds check.
An LLM may report these accesses as a bug or suggest adding a
length check. The reads are safe if \texttt{p} addresses four
readable bytes, but this precondition is not stated explicitly
in the source. Verification must therefore check whether its
callers establish it.  \tool does not guess that contract in one shot; the agent
refines its proposal by using counterexample feedback from the solver.
\begin{enumerate}[noitemsep, leftmargin=1.6em, label=\arabic*.]
  \item \emph{propose}: the specification agent initially proposes the precondition \texttt{p != NULL};
  \item \emph{check}: under this proposal, BMC analyses \texttt{read\_u32} in isolation and
        returns a counterexample trace in which a non-null~\texttt{p} addresses fewer than
        four readable bytes, causing an out-of-bounds read at \texttt{p[3]};
  \item \emph{validate}: the trace is genuine but unreachable from any observed
        in-tree call site. The trace shows no active defect and enters
        the refinement loop 
        rather than being reported as a crash;
  \item \emph{refine}: the contract refiner tightens the precondition to
        \texttt{valid\_range(p, 0, 4)}. To justify this refinement for the
        observed callers, their reachable inputs must imply the strengthened
        precondition; the formal obligation is given in \S\ref{sec:refinement};
  \item \emph{re-check}: BMC proves \texttt{read\_u32} safe under the refined
        contract, which then propagates to its callers as a stub.
\end{enumerate}
The LLM proposes a contract from the source and calling contexts.
The solver checks whether the function satisfies it within the
selected bounds. When a counterexample is found, the agent
examines the trace and its reachability from callers to
distinguish a defect from a missing precondition. The agent
may then propose a refinement, which must pass the soundness
guard before the function is checked again.

\begin{lstlisting}[
    float=tp,
    language=C,
    label=lst:packet,
    basicstyle=\ttfamily\footnotesize,
    abovecaptionskip=6pt,
    belowcaptionskip=4pt,
    aboveskip=6pt,
    caption={A packet parser with an implicit memory-safety contract. The helper \texttt{read\_u32} requires four readable bytes, but leaves this requirement to its callers to establish.}
]

uint32_t read_u32(const uint8_t *p) {
    return (uint32_t)p[0]
         | ((uint32_t)p[1] << 8)
         | ((uint32_t)p[2] << 16)
         | ((uint32_t)p[3] << 24);
}

static uint32_t 
decode_header(const uint8_t *buf, size_t len) {
    if (len < 8)
        return 0;

    uint32_t count = read_u32(buf + 4);
    if ((uint64_t)8 + (uint64_t)count * 4 > len)
        return 0;

    uint32_t sum = 0;
    for (uint32_t i = 0; i < count; i++) {
        uint64_t offset = 8 + (uint64_t)i * 4;
        sum += read_u32(buf + offset);
    }

    return sum;
}

uint32_t 
parse_packet(const uint8_t *buf, size_t len) {
    if (len < 8 || read_u32(buf) != PKT_MAGIC)
        return 0;

    return decode_header(buf, len);
}
\end{lstlisting}

\paragraph{Compositional verification}
Once \texttt{read\_u32} has been verified under its contract,
\tool uses that contract to verify its callers. Each call is
represented by a postcondition-constrained stub. Sound composition requires
establishing the helper's precondition at the call site. Under this obligation,
the stub allows
BMC to check the caller's safety and specified postconditions
without reanalysing the helper's implementation.
\texttt{parse\_packet} checks the header length, and
\texttt{decode\_header} checks that the declared payload fits
within \texttt{len}. Provided that \texttt{buf} contains at
least \texttt{len} readable bytes, these checks establish
the helper's precondition at the internal call sites. 

\paragraph{Contracts at the public boundary}
Both \texttt{read\_u32} and \texttt{parse\_packet} are exported
and rely on input contracts. The helper requires four readable
bytes. The parser requires a buffer containing at least
\texttt{len} readable bytes. Verification of internal callers
does not establish that external callers meet these requirements.
Passing a short buffer to the helper or overstating the buffer
length to the parser can cause an out-of-bounds read.
Whether such a contract violation constitutes a security defect
depends on the intended API contract and the reachability of
the offending call.

\paragraph{Role of \tool}
\tool coordinates contract inference, caller verification,
and counterexample validation. The agents propose contracts
and refinements using the source and calling contexts.
The backend checks the resulting obligations. This process
makes the conditions behind a verification result explicit
and identifies contract violations that require further
reachability analysis.
%
\section{\tool Architecture}
\label{sec:architecture}

\begin{figure*}[t]
\centering
\includegraphics[width=0.95\textwidth]{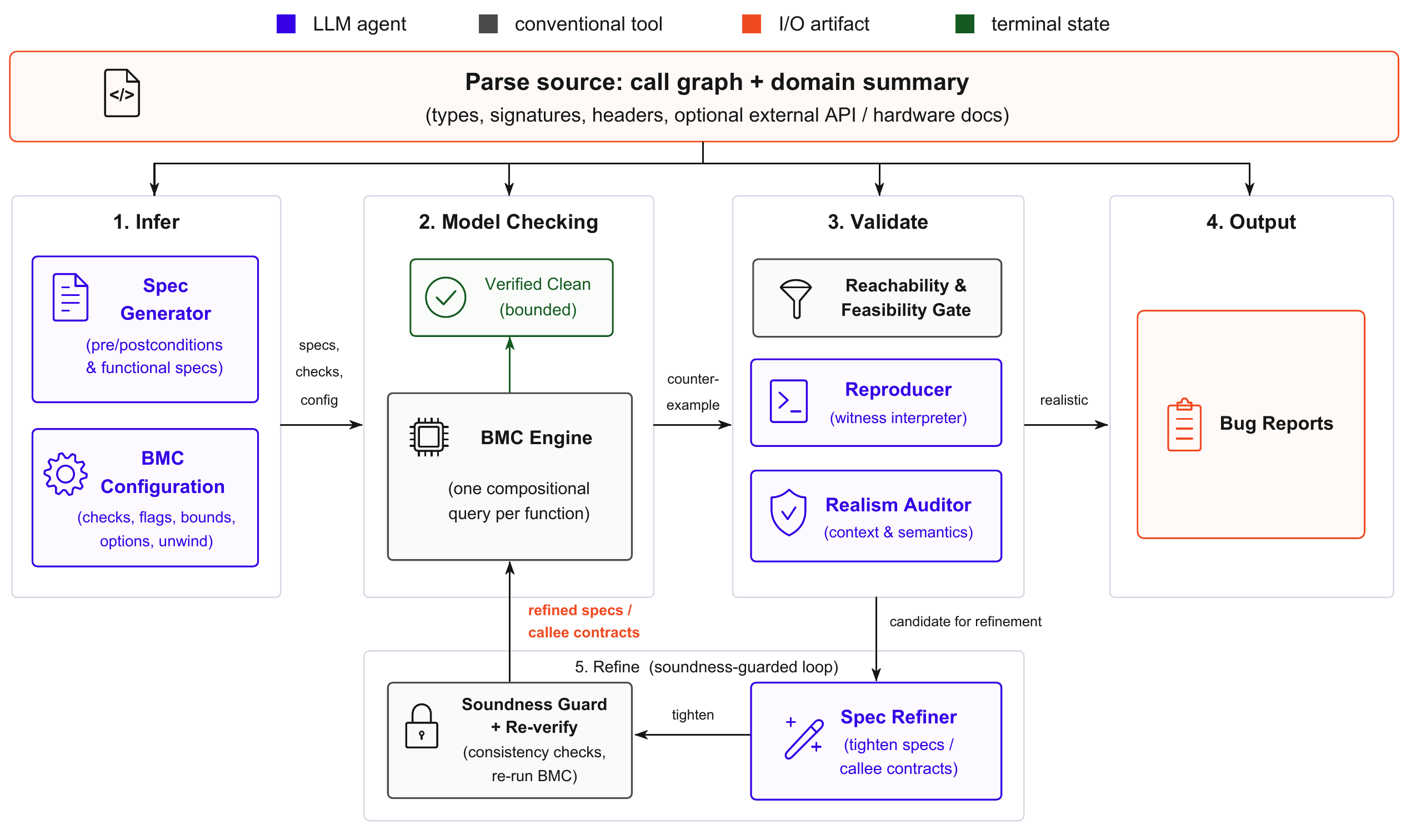}
\caption{BMC-Agent architecture}
\label{fig:architecture}
\end{figure*}

\begin{algorithm}[t]
\caption{\tool}
\label{alg:bmcagent}
\KwIn{source $S$}
\KwOut{findings $R$; contracts $C$}
$(F,G)\leftarrow\lit{Parse}(S)$\tcc*{call graph}
$C\leftarrow\lit{Infer}(F,G)$\nllabel{ln:infer}\tcc*{contracts, \S\ref{sec:specgen}}
$\theta\leftarrow\lit{Config}(F)$\nllabel{ln:config}\tcc*{configurations, \S\ref{sec:flagsel}}
$V\leftarrow\lit{BMC}(\lit{Harness}(F,C,\theta))$\nllabel{ln:harness}\tcc*{per function, \S\ref{sec:verify}}
$W\leftarrow\{(f,w)\mid w\in\lit{Cex}(V[f])\}$\nllabel{ln:work}\tcc*{witnesses to triage}
\ForEach{$(f,w)\in W$}{
  $e\leftarrow$\Validate{$f,w$}\nllabel{ln:validate}\tcc*{evidence for $w$}
  \lIf{$e$ confirms $w$}{$R\leftarrow R\cup\{(f,w,e)\}$}\nllabel{ln:report}
  \lElse{$W\leftarrow W\cup\Refine{f,w,e}$\nllabel{ln:refine}}
}
\Return $R,C$\;
\BlankLine
\Proc{\Validate{$f,w$}\tcc*[f]{\S\ref{sec:validation}}}{
  \lIf{$\lit{Feasible}(f,w)$ fails}{\Return \lit{infeasible}}
  \lIf{$\lit{Reach}(f,w)$ fails}{\Return \lit{unreachable}}
  $\mathit{run}\leftarrow\lit{Replay}(f,w)$\tcc*{from system entry}
  \Return $\lit{Audit}(f,w,\mathit{run})$\tcc*{realistic, or not}
}
\BlankLine
\Proc{\Refine{$f,w,e$}\tcc*[f]{\S\ref{sec:refinement}}}{
    \For{bounded iterations}{
    $\delta\leftarrow\lit{Propose}(f,w,C,e)$\tcc*{pre or callee post}
    \If{$\lit{Guard}(f,w,\delta)$ rejects}{\nllabel{ln:guard}
      add its refutation to $e$; \textbf{continue}\;
    }
    $C\leftarrow\lit{Apply}(C,\delta)$\; 
    $V'\leftarrow\lit{BMC}(\lit{Harness}(f,C,\theta))$\nllabel{ln:recheck}\;
    {\Return $\{f\}\times \lit{Cex}(V')$}\nllabel{ln:back}
  }
  record $w$ as unresolved; \Return $\emptyset$\;
}
\end{algorithm}

\subsection{Overview}

Figure~\ref{fig:architecture} shows how \tool orchestrates an off-the-shelf bounded model checker, and Algorithm~\ref{alg:bmcagent} outlines it in
pseudocode.  The system
parses the source,
builds the call graph, and summarises domain information from types,
signatures, headers, and optional documentation. \tool infers a contract for every function (line~\ref{ln:infer}), configures and checks each function (lines~\ref{ln:config} and~\ref{ln:harness}), validates the
counterexample before reporting it (line~\ref{ln:validate}), and refines a
contract when validation rejects a witness (line~\ref{ln:refine}).
The five stages of
Figure~\ref{fig:architecture} are lines~\ref{ln:infer}
and~\ref{ln:config}, line~\ref{ln:harness}, \lit{Validate},
line~\ref{ln:report}, and \lit{Refine}. The remainder of this section
takes them in that order.


\paragraph*{Agents propose, solvers verify}
No agent proposal is accepted on its own authority. A contract, a configuration, a
reproducer, and a refinement are all candidates that some check must
accept. The backend decides whether a counterexample exists, compilation
and execution decide whether a witness reproduces, and the refinement guard
decides whether a revision may be applied. A successful result applies
within the model and bounds of its query, and each bug report records the
evidence obtained during validation. Table~\ref{tab:agents} summarises the
agents and the checks applied to their proposals.


\begin{table*}[t]
\centering
\caption{The LLM agents in \tool. Agents propose specifications,
configurations, and refinements. The BMC engine (\S\ref{sec:bmc})
checks the encoded properties, and validation assesses the resulting
witnesses before reporting findings.}
\label{tab:agents}
\small
\renewcommand{\arraystretch}{1.4} 
\setlength{\aboverulesep}{0pt}    
\setlength{\belowrulesep}{0pt} 
\begin{tabularx}{\textwidth}{@{}p{0.18\textwidth}p{0.08\textwidth}X X@{}}
\toprule
\textbf{Agent} &
\textbf{Stage} &
\textbf{Proposal} &
\textbf{Checks on agent outputs} \\
\midrule

Spec Generator (\S\ref{sec:specgen})
&
infer
&
Function body, caller context, domain summary
$\rightarrow$
pre/postconditions and optional functional specifications
&
BMC checks the function
against the encoded clauses under the proposed specification.
\\

\makebox{BMC Configuration} (\S\ref{sec:flagsel})
&
infer
&
Function and callee bodies, call graph $\rightarrow$ checks, bounds, and timeouts, and per-callee inline-versus-abstract (scope)
&
The agent cannot remove a callee from the statically inlined set, and a failed proposal falls back to the default configuration.
\\

Reproducer (\S\ref{sec:validation})
&
validate
&
Witness, call chain, source context
$\rightarrow$
a system-entry harness driving the witness inputs from the highest available caller down to the fault site (entry function, includes, linkage)
&
Compilation and execution determine whether the generated harness
builds and produces a runtime fault.
\\

Realism Auditor (\S\ref{sec:stage4})
&
validate
&
Specification, harness, witness state, source context
$\rightarrow$
realistic or unrealistic assessment
&
Deterministic detectors reject known artifact classes before the LLM is consulted. The remaining assessments are recorded as
evidence, and any refinement they trigger is subject to the refinement
guard.
\\

Spec Refiner (\S\ref{sec:refinement})
&
refine
&
Witness, harness, source
$\rightarrow$
tightened preconditions or callee contracts
&
The refinement guard tests the revision against
Definition~\ref{def:admissible}, and the function is checked again under
the revised contracts.
\\
\bottomrule
\end{tabularx}
\end{table*}

\subsection{Specification Inference and BMC Configuration}
\label{sec:propose}

Two agents prepare each function for verification.  The specification
generator proposes a contract describing the inputs the function requires
and the guarantees it provides (\S\ref{sec:specgen}).  The BMC configuration
agent decides which properties to check, which bounds and timeouts to
use, and which callee bodies to include in the query (\S\ref{sec:flagsel}).

\subsubsection{Specification Generation}
\label{sec:specgen}
A function contract $C_f=(P_f,Q_f)$ consists of a precondition $P_f$
and a postcondition $Q_f$. The precondition states the conditions under which the function may
be called. The postcondition states what the caller can rely on
when the function returns normally. We write contracts in the following notation.
\[
\begin{aligned}
C_f &::= \texttt{pre:}\;\varphi_{\mathrm{pre}}
       \quad \texttt{post:}\;\varphi_{\mathrm{post}} \\
\varphi &::= b \mid e_1 \bowtie e_2 \mid \pi(e_1,\ldots,e_n) \\
        &\quad \mid \neg\varphi \mid \varphi\land\varphi
               \mid \varphi\lor\varphi .
\end{aligned}
\]
Here $b$ is a Boolean constant or Boolean-valued program expression,
$\bowtie$ is a comparison operator, and $\pi$ is a predicate. Each $e$ is a program expression and
may include arithmetic, bitwise operations, indexing, or field access.
Memory predicates include \lstinline{valid(p)},
\lstinline{in_bounds(a,i)}, and \lstinline{valid_range(p,lo,hi)}.
Within the selected bounds, BMC assumes
the encoded precondition at entry, checks the selected safety properties
during execution, and asserts the encoded postcondition on normal return.

\smallskip\noindent\textbf{Caller-grounded inference.}
\label{sec:functional-spec}
A contract inferred from the function body alone can mistake an
implementation error for intended behaviour. The caller-grounded
generator therefore drafts the contract from the body and a bounded number of 
observed call sites, giving priority to conditions supported by
the call sites.
The body shows
which conditions its operations require. The call sites show which
arguments callers pass, what they check before the call, and how they
use the returned value. 

The generator processes callees before callers
(line~\ref{ln:infer} of Algorithm~\ref{alg:bmcagent}) and those callee
contracts exist when a caller is drafted. Inference completes for every function
before the first verification query runs. Counterexample-driven refinement adds clauses to
these contracts rather than redrafting them
(\S\ref{sec:refinement}).

\smallskip\noindent\textbf{Revisiting the contract.}
The generator records a disagreement when the body and the observed calls
suggest conflicting preconditions or postconditions.  The agent can retrieve
additional call sites, inspect callees and structure definitions, and revise
the contract.  

\subsubsection{BMC Configuration}
\label{sec:flagsel}

For each function, the configuration agent reads its body and can
retrieve callee bodies as needed.  It selects the optional backend
checks, unwinding bounds, and per-function timeout. Optional checks include
signed- and unsigned-integer overflow, lossy type conversions, and
out-of-bounds pointer arithmetic. Enabling every check on every function would bury real violations in noise, so the agent enables a check only where the function’s code makes it meaningful. These choices determine which
properties the query evaluates and within which bounds.

The agent also selects which callee bodies to inline. An inlined
callee is analysed directly. An abstracted callee is represented by
a postcondition-constrained stub. This choice determines
the scope of the query and which callees the checker analyses through
their bodies or contracts. 
A static rule inlines small callees that are file-local and free of loops, allocation, recursion, and function-pointer dispatch. The agent then reads the remaining bodies and inlines those a contract would approximate too loosely, such as small accessors whose stub would return an arbitrary value, leaving the rest abstracted.


\subsection{Per-Function Model Checking with Compositional Stubs}
\label{sec:verify}

Given an inferred specification and the per-function configuration, \tool constructs a
bounded model checking query for each function. It inlines selected callees and represents the remaining calls with stubs. Counterexamples are deduplicated before reachability validation.

\subsubsection{Verification Harness Construction}
\label{sec:bmc}

Harness construction is deterministic. Each harness initialises
nondeterministic inputs, adds environmental assumptions and applicable
invariants from the verification knowledge base, and translates the
specification into backend assumptions and assertions. It then invokes
the target function and checks its postcondition alongside the selected
safety properties.

Each query runs under a configurable unwinding bound $k$ and a
per-function timeout. When the backend returns an error rather than a
verdict, from a resource limit or an encoding failure, \tool retries with an adjusted configuration.
A~retry may also abstract a callee that was previously inlined.
%

\tool records three outcomes, \emph{verified clean},
\emph{counterexample}, and \emph{inconclusive}. A verified-clean
result records bounded verification relative to the encoded
specification, assumptions, and abstractions. A counterexample
proceeds to deduplication and validation. Unresolved timeouts,
encoding failures, and backend errors produce an inconclusive
result. A~vacuity check also treats backend success with zero
checked conditions as inconclusive.

\subsubsection{Counterexample Deduplication}
\label{sec:dedup}

A property violation may produce several witnesses along different
execution paths. Before validation, \tool groups counterexamples
by \texttt{(function, property type)} and selects representative
witnesses. This reduces repeated validation of similar failures.
The grouping is coarse and may combine distinct defects within
the same function and property class.

\subsection{Counterexample Validation}
\label{sec:validation}
  
A BMC counterexample is not immediately reported as a bug.  For real code at scale, more of the difficulty lies here than in the model checking.
Whether a function is genuinely buggy and whether a
counterexample is real depend on how the function is reached and
constrained by context, rather than on the function in isolation.  A trace that
violates a property in a standalone harness may be unreachable, or require
inputs that no caller can supply.  After
deduplication, each resulting witness therefore passes through the following
validation stages. 

  \paragraph*{Stage~1 -- Callee feasibility}
  A BMC witness is computed with callees replaced by stubs, so it may depend
  on an overapproximated return value.  Stage~1 substitutes the real callee
  bodies, fixes the inputs to the values in the trace, and re-runs the
  backend. A witness that survives does not depend on an over-permissive stub. Whether its inputs can arise at all is a separate question, settled in Stage~2.

  \paragraph*{Stage~2 -- Input reachability}
  \tool then propagates the surviving witness's inputs upward through the call
  graph, asking at each step whether the caller can supply the values the
  backend chose. Each bounded symbolic query checks one
  caller--callee edge.  The harness supplies nondeterministic caller
  inputs, stubs the other callees, and asserts that the target's arguments
  cannot match the witness. A counterexample to that assertion exhibits
  a caller that supplies them.  
  Edges are checked independently against the same witness, 
  and a chain of feasible edges shows
  that each step is possible rather than that one execution performs all
  of them.

  \paragraph*{Stage~3 -- Dynamic replay}
  For the surviving witnesses, the \emph{Reproducer} agent (Table~\ref{tab:agents})
  writes a program that calls the entry point at
  the top of the chain accepted in Stage~2 and drives it down to the fault
  site, then compiles and executes it. Observing the fault at runtime
  upgrades the finding to dynamically confirmed. Because the call chain
  is kept intact, a successful run exercises in one execution the path
  Stage~2 assembled an edge at a time. 
  
  The LLM writes the harness, but does not decide the outcome.  A captured signal
  counts only when it fires inside or after the call to the target
  function. \tool records the fault site and discounts signals raised in
  harness setup.  
  %
  A captured fault does not by itself establish that the entry sequence the program uses is one the system permits. Because the harness controls only part of the system state, real bugs
  that surface on the deployed platform may be missed.

  \paragraph*{Stage~4 -- Realism audit}
  \label{sec:stage4}
  Stages~1--3 establish that the witness is reachable in the model
  (symbolically, in Stages~1--2) and reproducible on a host harness
  (concretely, in Stage~3).  Neither guarantees that the witness could arise in the deployed system, because of reasons outside the BMC model and the target
  host harness. Examples are uninitialised framework globals, impossible hardware
  states, and path-divergent loop unwinding artifacts.  A final realism audit
  reads the specification, harness, witness state, and surrounding source
  code, and classifies the witness as \textbf{realistic} or
  \textbf{unrealistic}.

\subsection{Closing the Loop: Refinement and Re-verification}
\label{sec:loop}
\label{sec:refinement}

  Validation can identify witnesses that depend on infeasible callee
behaviour, unreachable inputs, or unrealistic environmental assumptions.
These witnesses are candidates for refinement. 

  \smallskip\noindent\textbf{Refinement proposal.}
  Given the witness, harness, and source context, an agent proposes one of
  two kinds of revision.
  \begin{itemize}[noitemsep]

    \item \emph{Function precondition.}
          Strengthen the precondition of the function under check to exclude the spurious witness that was rejected by validation.
    \item \emph{Callee contract.}  Tighten
          a stubbed callee's contract to exclude behaviour that its
        implementation cannot produce.
        
  \end{itemize}
  \smallskip\noindent\textbf{Admissible refinement.}
A counterexample can be removed by tightening the precondition
to exclude the witness state, and the
disappearance of a witness carries no information on its own.
A~refinement that records a condition the callers already enforce
recovers an implicit contract. A refinement that excludes states the
callers can produce deletes executions from the analysis. The model checker
cannot tell these apart, because the callers are not
part of its query. The guard therefore
restricts which clauses may be accepted.

\begin{definition}[Admissible refinement]
\label{def:admissible}
Let $w$ be a rejected witness of the query for $f$ under contracts $C$,
with $\sigma_0$ its entry state and callee returns. A refinement
$\delta$ adds one clause, $\Delta$ to the precondition of $f$ or
$\Gamma$ to the postcondition of a stubbed callee $g$, leaving the rest
of the query unchanged. It is \emph{admissible} if it is
\begin{itemize}[noitemsep]
  \item \emph{effective}: the strengthened contract excludes $\sigma_0$;
  \item \emph{satisfiable}: the strengthened contract admits an entry
        state of $f$, respectively a return of $g$;
  \item \emph{non-deleting}: every other witness of the query either
        satisfies the strengthened contract or is itself classified
        spurious by \S\ref{sec:validation};
  \item \emph{justified}: no observed caller reaches a call to $f$ in an
        entry state violating $\Delta$, respectively no execution of $g$
        from a state satisfying its precondition violates $\Gamma$.
\end{itemize}
\end{definition}

\begin{lemma}[Guarantees of admissible refinement]
\label{lem:admissible}
Call a witness \emph{realizable} when an observed caller can supply its
entry state and the real callee can produce its callee returns from the
state at which it is called. An admissible refinement retains every
realizable witness of the query, and removes no other witness without a
verdict from \S\ref{sec:validation}.
\end{lemma}
\begin{IEEEproof}
For a precondition revision, justification ensures that the entry state
of every realizable witness satisfies $\Delta$. For a
callee-postcondition revision, the call site establishes $g$'s
precondition, and justification then ensures that every real callee
return in such a witness satisfies $\Gamma$. Thus every realizable
witness satisfies the added clause, and as a witness of the query it
already satisfies the unstrengthened contract, hence the strengthened
one. Since the remainder of the query is unchanged, its trace remains admitted and violates the same property.
The non-deleting condition directly ensures that every other excluded
witness has been classified spurious. 
\end{IEEEproof}

  \smallskip\noindent\textbf{What \tool enforces.}
The guard (line~\ref{ln:guard} of Algorithm~\ref{alg:bmcagent})
approximates Definition~\ref{def:admissible}. \emph{Effectiveness} holds
by construction. \emph{Satisfiability} reduces to reachability on the assumptions and can be checked with an \lstinline{assert(0)}
probe. The guard leaves
\emph{non-deletion} open, as the condition ranges over witnesses the
query never enumerated, and an accepted clause may exclude a witness BMC never reported. The retention guarantee of Lemma~\ref{lem:admissible}
survives this, following from justification alone.
\emph{Justification} rests on a solver check that asserts the clause at
the observed call sites and the callee return. 

  \smallskip\noindent\textbf{Re-verification.}
An accepted revision is applied to the affected contracts, and $f$ is
checked again (line~\ref{ln:recheck}) under the revised contracts. Its
counterexamples return to triage (line~\ref{ln:back}), each receiving
its own verdict. The clause removes the rejected witness. Another
state can violate the same property, and that witness comes back
through the work list. Witnesses of $f$ still queued are screened
against the revised precondition.

\subsection{Scaling BMC through Orchestration}
\label{sec:scalability}

The orchestration above is what lets \tool drive an off-the-shelf bounded model
checker at the scale of a large, real-world codebase. 
In particular, \tool introduces five mechanisms for scalability. 

\smallskip\noindent\textbf{Per-function decomposition using assume-guarantee.}
Each function is checked against its specification with selected callees
replaced by postcondition stubs (\S\ref{sec:bmc}). A verified result
transfers to the caller with real callee bodies only if each call meets
the callee's precondition, each abstracted callee satisfies its contract
and safety checks, and the stubs cover its concrete effects. This transfer
also requires a faithful bounded encoding and callee checks that cover
the executions admitted at each call site.

\smallskip\noindent\textbf{Parallel dispatch.}
Per-function queries are dispatched concurrently using the available
specifications (\S\ref{sec:specgen}), with check selection
(\S\ref{sec:flagsel}) running as a parallel pre-pass. Call-graph information
supports caller grounding and identifies the callers a changed contract
affects. 

\smallskip\noindent\textbf{Witness deduplication before validation.}
A single defect often yields many related counterexamples; \tool merges them
before validation (\S\ref{sec:dedup}), and keeps a few representatives per
property class. The validation then runs a handful of times per defect rather than once per witness.

\smallskip\noindent\textbf{Agentic BMC configuration.}
Per-function configuration selects checks, bounds, timeouts, and
whether callees are inlined or abstracted (\S\ref{sec:flagsel}). This adapts
each query to its target property and resources; the guarantees remain
relative to those choices and the abstraction premises.

\smallskip\noindent\textbf{Incremental persistence.}
A specification store persists tightened specifications and reloads them on re-verification and
resume, which spares refinement and repeated verification runs from regenerating
specifications for functions already analysed; a pattern library short-circuits the realism audit
(\S\ref{sec:stage4}) on recurring artifacts before any LLM call.


\section{Evaluation}
\label{sec:evalsummary}

We evaluate \tool's agentic orchestration through four research questions: 

\begin{description}
\item[\textbf{RQ1: Comparison with traditional verifiers.}]
%
How does \tool compare with established automatic verifiers such as CBMC~\cite{cbmc},
ESBMC~\cite{cordeiro2011smt}, and CPAchecker~\cite{beyer2011cpachecker} on standard verification benchmarks?

\item[\textbf{RQ2: Comparison with agentic verification.}]
Does \tool discover more confirmed defects than existing agentic
verification systems on realistic codebases?

\item[\textbf{RQ3: Specification quality.}]
Are the specifications generated by \tool competitive, as standalone
artifacts, with those generated by specialised synthesizers?

\item[\textbf{RQ4: Real-world applicability.}]
Can \tool produce actionable findings on mature open-source software and
heavily fuzzed codebases?

\end{description}

\subsection{Experimental Setup}
\label{sec:setup}

\tool uses Claude Sonnet~4.5/4.6 as its agent backbone, CBMC~5.95.1
(default) and ESBMC~8.3.0 as the C verification backends, and Kani~0.67.0 as
the Rust verification backend.


We evaluate on four corpora spanning both generated and human-written software:
\begin{itemize}

\item \textbf{\vibeos}~\cite{vibeos}: a 15\,KLoC ARM64 operating system
written in C with substantial LLM assistance, based on commit
\texttt{866bfb9b7fc5}.

\item \textbf{Claude's C Compiler (CCC)}~\cite{ccc}: a C compiler
comprising approximately 50\,KLoC of Rust, generated primarily
through LLM-assisted development, based on commit \texttt{6f1b99acb2f4}.

\item \textbf{OSS-Fuzz and mature open-source software}~\cite{chang2016oss}:
jq, OpenSSL, libcurl, libxml2, protobuf-upb, and additional
manually selected targets.

\item \textbf{SV-COMP 2026 benchmarks}~\cite{beyer2026evaluating}: a subset
that covers memory safety, arithmetic overflow, and reachability properties.

\end{itemize}


\subsection{RQ1: Comparison with Traditional Verification Tools}
\label{sec:rq5-tools}


\paragraph{Benchmarks}
We evaluate on two SV-COMP sets, both for the \texttt{unreach-call} property
(reachability of \texttt{reach\_error}). 
\textbf{Tier~A:} \textsc{aws-c-common}, derived from the Amazon Web Services (AWS) C Common
utility library (all 7 bug tasks plus 20 safe tasks). 
\textbf{Tier~B:} all 29 tasks of
\textsc{ldv-linux-3.4-simple} labelled \emph{false} for
\texttt{unreach-call}. 
These are
Linux device-driver harnesses of 0.18--1.74\,MB of CIL-preprocessed C spanning
many functions. 
Tier~B is thus a \emph{falsification} experiment, scored against targets confirmed reachable. 

\paragraph{Baselines} 
We compare \tool against the three established verifiers summarised in
Table~\ref{tab:rq-baselines}. For \tool
a catch is a solver-confirmed target; for the standalone verifiers
it is a reported target violation.

\begin{table}[th]
\centering
\caption{Baselines for comparison to answer RQ1}
\label{tab:rq-baselines}
\normalsize
\setlength{\tabcolsep}{4pt}\renewcommand{\arraystretch}{1.1}
\noindent\begin{tabularx}{\columnwidth}{@{}lX@{}}
\toprule
\textbf{Verifier} & \textbf{Technique and configuration} \\
\midrule
CBMC & Bounded model checking; \tool's default BMC backend. \\
ESBMC & SMT-based model checking; bounded mode as a \tool backend and $k$-induction for the standalone baseline. \\
CPAchecker & Configurable program analysis with CEGAR-based predicate abstraction; SV-COMP configuration. \\
\bottomrule
\end{tabularx}
\end{table}


\begin{table}[!htb]
\centering
\small
\setlength{\tabcolsep}{3pt}\renewcommand{\arraystretch}{1.1}
\caption{Tier~A control: \textsc{aws-c-common}, 27-task sample (20 safe / 7 bug),
\texttt{unreach-call} property, LP64 data model. \tool's CBMC run uses
unwinding bound~64; its ESBMC run uses bounded mode. The standalone ESBMC
baseline uses $k$-induction. 
}
\label{tab:rq5-aws}
\begin{tabularx}{\columnwidth}{@{}>{\raggedright\arraybackslash}Xcccc@{}}
\toprule
\textbf{Tool} & \textbf{\shortstack{Safe\\proved}} & \textbf{\shortstack{False\\alarms}} & \textbf{\shortstack{Bugs\\caught}} & \textbf{Unknown}\\
\midrule
\tool + CBMC         & \textbf{18/20} & \textbf{0} & \textbf{7/7} & \textbf{2}\\
\tool + ESBMC        & 13/20 & 2 & 6/7 & 6\\
CBMC                 & \textbf{18/20} & \textbf{0} & \textbf{7/7} & \textbf{2}\\
ESBMC ($k$-induction) & 13/20 & 3 & 5/7 & 6\\
CPAchecker           & 5/20  & 1 & 4/7 & 17\\
\bottomrule
\end{tabularx}
\end{table}

\paragraph{Tier~A -- \tool matches the baseline where BMC is tractable}
On \textsc{aws-c-common}, \tool with CBMC proves $18/20$ safe tasks and
catches all $7/7$ bugs with no false alarms. 
CBMC also scores $18/20$ and $7/7$, and the two leave the
{identical} pair of safe tasks reported as unknown (both SAT-capacity
limits).  On these harnesses, \tool's orchestration inlines the
full call closure, issuing effectively the same monolithic query as CBMC. 
The agentic orchestration shows no regression on
code BMC already handles.  With ESBMC as its backend, \tool proves $13/20$
safe tasks and catches $6/7$ bugs with two false alarms.  This preserves the
standalone ESBMC baseline's safe-proof coverage, while improving bug catching
from $5/7$ to $6/7$ and reducing false alarms from three to two; its remaining
unknowns are timeouts or out-of-memory outcomes.  CPAchecker leaves most
tasks unknown (Table~\ref{tab:rq5-aws}).

\paragraph{\tool improves bug-finding on Linux drivers}
Every task in this set is expected-unsafe, with the target reachable in all 29, and a tool is scored on whether it finds a path to it. Standalone CBMC reaches $10/29$ targets. 
Using the \emph{same} CBMC engine, \tool reaches it in
$18/29$ tasks, including all ten reached by standalone CBMC. With ESBMC as
its backend in bounded mode, \tool confirms reachability in $15/29$
tasks. CPAchecker reaches $13/29$ targets and standalone ESBMC reaches
$10/29$ and returns one incorrect safe verdict (\texttt{megaraid\_mm}). 
\tool with
CBMC reaches the most targets 
(Table~\ref{tab:rq5-ldv}).

\begin{table}[t]
\centering
\setlength{\tabcolsep}{4pt}\renewcommand{\arraystretch}{1.15}
\caption{Tier~B frontier: 29 \textsc{ldv-linux-3.4-simple} tasks under ILP32.
\emph{Catches} counts solver-confirmed targets for \tool and reported target
violations for standalone verifiers. \emph{Unsound} counts safe verdicts
that conflict with confirmed target reachability under ILP32.
\emph{Unknown} counts the remaining tasks.}
\label{tab:rq5-ldv}
\normalsize
\begin{tabular}{@{}lccc@{}}
\toprule
\textbf{Tool} & \textbf{Catches} & \textbf{Unsound} & \textbf{Unknown}\\
\midrule
\tool + CBMC         & 18/29 & 0 & 11\\
\tool + ESBMC        & 15/29 & 0 & 14\\
CPAchecker           & 13/29 & 0 & 16\\
ESBMC (k-induction)  & 10/29 & \textbf{1} & 18\\
CBMC                 & 10/29 & 0 & 19\\
\bottomrule
\end{tabular}
\end{table}

Of \tool's 18 CBMC-backed confirmations, four are reached by neither
standalone ESBMC nor CPAchecker. Each of those two tools also reaches
targets that the CBMC-backed \tool run leaves unconfirmed. The tools thus
complement one another on these driver harnesses.


\subsection{RQ2: Comparison with FM-Agent}
\label{sec:rq2-fmagent}

RQ2 compares \tool with \fmagent~\cite{fmagent}, the closest prior LLM-based
verification agent.  We run both systems on the same \vibeos kernel corpus and
manually review their outputs, counting a finding only when the reviewed source
contains a concrete bug at a reachable \vibeos boundary; reports that require
additional backend, environment, or helper-level assumptions are excluded.  We
do not use target reproduction as a comparison column, because the current
\fmagent artifact provides no equivalent target-replay layer.

\begin{table}[!htb]
\centering
\setlength{\tabcolsep}{3pt}\renewcommand{\arraystretch}{1.12}
\caption{RQ2: \tool vs.\ \fmagent on the \vibeos kernel corpus.  Raw candidates
are each tool's native output unit; confirmed bugs are manually source-validated
reports.  Confirmed functions deduplicate those reports by function name.}
\label{tab:rq2-fmagent-vibeos-summary}
\normalsize
\begin{tabular}{@{}lrrrr@{}}
\toprule
\textbf{Tool} &
\textbf{\shortstack{Raw\\ cand.}} &
\textbf{\shortstack{Confirmed\\ bugs}} &
\textbf{\shortstack{Confirmed\\ /raw}} &
\textbf{\shortstack{Confirmed\\ functions}} \\
\midrule
\fmagent & 120 & 20 & 17\% & 20 \\
\tool    & 43  & 21 & 49\% & 13 \\
\bottomrule
\end{tabular}
\end{table}

\paragraph{Results}
Table~\ref{tab:rq2-fmagent-vibeos-summary} summarises the results.  \fmagent
reports at the function level: its 120 candidate mismatches yield 20
confirmed bugs in 20 distinct functions.  \tool reports at the finer
checked-property level: its 43 candidates yield 21 confirmed bugs in 13 distinct
functions, and one function can contribute several confirmed bugs when independent
properties fail.

\begin{table*}[t]
\centering
\setlength{\tabcolsep}{5pt}\renewcommand{\arraystretch}{1.12}
\caption{Confirmed \vibeos findings by primary bug class. Entries give
report-level confirmed bug counts and distinct function counts, followed by the function names where applicable. A
function can appear in more than one row when different confirmed
properties are in that function.}
\label{tab:rq2-vibeos-confirmed-functions}
\normalsize
\begin{tabularx}{\textwidth}{@{}p{0.16\textwidth}XX@{}}
\toprule
\textbf{Class} & \textbf{\fmagent} & \textbf{\tool} \\
\midrule
\emph{Numeric/range} &
\begin{minipage}[t]{\linewidth}\raggedright
\textbf{11 bugs / 11 funcs:}
\texttt{fat32\_file\_size}, \texttt{fat32\_init},
\texttt{fat32\_read\_file\_offset}, \texttt{read32},
\texttt{do\_request}, \texttt{vfs\_read}, \texttt{mouse\_set\_pos},
\texttt{calloc}, \texttt{malloc}, \texttt{fb\_draw\_char},
\texttt{fb\_fill\_rect}.
\end{minipage}
&
\begin{minipage}[t]{\linewidth}\raggedright
\textbf{13 bugs / 9 funcs:}
\texttt{elf\_calc\_size}, \texttt{elf\_process\_relocations},
\texttt{elf\_load\_at},
\texttt{fb\_draw\_char}, \texttt{fb\_draw\_string},
\texttt{malloc}, \texttt{dns\_resolve}, \texttt{vfs\_write},
\texttt{vfs\_append}.
\end{minipage} \\
\emph{Object-bound/pointer} &
\begin{minipage}[t]{\linewidth}\raggedright
\textbf{2 bugs / 2 funcs:}
\texttt{vfs\_readdir}, \texttt{vfs\_lookup}.
\end{minipage}
&
\begin{minipage}[t]{\linewidth}\raggedright
\textbf{7 bugs / 6 funcs:}
\texttt{elf\_process\_relocations}, \texttt{elf\_load\_at},
\texttt{fb\_draw\_char}, \texttt{process\_create},
\texttt{readdir\_callback}, \texttt{vfs\_readdir}.
\end{minipage} \\
\emph{API contract/}\newline
\emph{ordering} &
\begin{minipage}[t]{\linewidth}\raggedright
\textbf{6 bugs / 6 funcs:}
\texttt{virtio\_sound\_play\_pcm}, \texttt{vfs\_rename},
\texttt{vfs\_mkdir}, \texttt{vfs\_create},
\texttt{icmp\_send\_echo\_request}, \texttt{kapi\_delete\_dir}.
\end{minipage}
&
\textbf{0} \\
\emph{Protocol/format} &
\begin{minipage}[t]{\linewidth}\raggedright
\textbf{1 bug / 1 func:}
\texttt{ip\_handle}.
\end{minipage}
&
\begin{minipage}[t]{\linewidth}\raggedright
\textbf{1 bug / 1 func:}
\texttt{ip\_handle}.
\end{minipage} \\
\bottomrule
\end{tabularx}
\end{table*}

Table~\ref{tab:rq2-vibeos-confirmed-functions} breaks the
confirmed findings down by bug class.
We use four bug classes. \emph{Numeric/range} covers integer overflow,
truncation, or signed-shift undefined behaviour (e.g.\ \texttt{read32} and
\texttt{fat32\_init} reconstruct a 32-bit field with \texttt{p[3]<<24}, shifting
a high byte into the sign bit). \emph{Object-bound/pointer} covers out-of-bounds
reads or writes and invalid pointer arithmetic (e.g.\ \texttt{elf\_process\_relocations}
writes the relocated value past the mapped target). \emph{API contract/ordering} covers
violations of documented contracts, such as a partial failure reported as success
(\texttt{virtio\_sound\_play\_pcm} returns \texttt{0} after a mid-buffer submit
fails) or validation performed after an unsafe operation. 
\emph{Protocol/format} covers a packet length field used without a bounds check
(\texttt{ip\_handle} underflows the IPv4 payload length).

\paragraph{Discussion}
The two systems are complementary rather than one dominating the other.  \fmagent
produces broader, specification-oriented reports, more evenly distributed across
classes (notably the API-contract and ordering defects, none of which \tool confirms on \vibeos).  \tool produces fewer candidates but grounds each in a CBMC
counterexample, concentrating on implementation-level memory-safety and
undefined-behaviour defects (out-of-bounds accesses, invalid pointer arithmetic,
integer overflow) at untrusted-input boundaries such as ELF, packet, and
filesystem parsing, where unconstrained inputs are the realistic threat model and
the counterexample trace is a concrete witness of reachability.  Their
confirmed-function overlap is small but non-empty: \texttt{malloc},
\texttt{fb\_draw\_char}, \texttt{vfs\_readdir}, and \texttt{ip\_handle}. 

\paragraph{Intermediate statistics}
We report intermediate counts from five stages of BMC-Agent's analysis of 600 functions. During harness construction, 
546 of 600 (91.0\%) harnesses 
compile without repair; the agent
rebuilds 53 of 54 failed harnesses, yielding 599 of 600 (99.8\%). Validation
passes 161 distinct (function, property) pairs to the realism audit
after deduplication merges 3,670 raw witness records into 785 groups. 
Across 611 refinement iterations, 
95 tightenings are rejected. 
Finally, source review confirms 21 of the 43 candidates that reach final
reporting.

\paragraph{Cost and runtime}
Across the 25 \vibeos module runs used for RQ2 (600 
functions), the summed run duration was 14.1\,h, averaging 84.6 seconds
per function. Usage was logged for 1{,}107 of 2{,}586 agent calls (42.8\%)
and implies US\$38.40 for those calls. Extrapolating to all calls under an
equal-average-cost assumption gives an estimated total cost of US\$89.71,
or about US\$0.150 per function.

\paragraph{Component ablation}
We ablate three components of \tool on 47 functions from four \vibeos
modules (\texttt{elf}, \texttt{fb}, \texttt{memory}, and \texttt{string}).
These functions contain 11 of the 21 confirmed findings in
Table~\ref{tab:rq2-vibeos-confirmed-functions}; we use these findings as
the reference set. Table~\ref{tab:vibeos-ablation} compares the Full
configuration with three ablated configurations on the same functions.

\begin{table}[t]
\centering
\small
\setlength{\tabcolsep}{4pt}
\caption{Component ablation on four \vibeos modules (47 functions). Raw candidates are the tool’s native output
unit. The Confirmed column counts distinct findings from
Table~\ref{tab:rq2-vibeos-confirmed-functions}.}
\label{tab:vibeos-ablation}
\begin{tabular}{@{}lrrr@{}}
\toprule
\textbf{Configuration}  &
\textbf{Raw candidates} & \textbf{Confirmed} \\
\midrule
Full          & 26 & 11 \\
No dynamic    & 18 &  7 \\
No realism    & 11 &  1 \\
Fixed BMC config  &  6 &  2 \\
\bottomrule
\end{tabular}
\end{table}

Dynamic replay tests whether CBMC counterexamples execute in a compiled harness: removing it reduces candidates from 26 to 18 and confirmed findings from 11 to 7. The realism audit both qualifies candidates and supplies the semantic signal for specification repair; without it, candidates fall to 11 and confirmed findings to one.  Fixed BMC configuration affects an earlier stage: without the configuration agent, flag selection, and inlining advisor,  candidates fall to 6 and
confirmed findings to 2.

\subsection{RQ3: Specification Quality}
\label{sec:rq2}

To assess whether inferred specifications support deductive verification,
we run \tool in specification-emitting mode on the C benchmarks from
AutoSpec~\cite{autospec} and the Java benchmarks from SpecGen~\cite{ma2025specgen}.
We report the number of cases for which the verifier discharges all
obligations in at least one attempt and the fraction of all attempts in
which it does so.

\paragraph{C/ACSL (AutoSpec)}
\tool and AutoSpec each make five independent attempts per C program
using Claude Sonnet~4.6. 
Frama-C/WP checks the generated ACSL annotations and the program properties
supplied by the benchmark. \tool solves 209 of 251 programs, compared
with 219 in our AutoSpec rerun and 199 in AutoSpec's published evaluation.
AutoSpec's lead in the rerun comes from the Frama-C-problems, SV-COMP,
and OOPSLA subsets, where its bottom-up loop refinement retries inference
of missing invariants. On SyGuS/Code2Inv, each system solves 121 of
133 cases, and \tool has the higher per-attempt success rate.

\commentout{
For example, Code2Inv case \texttt{10.c} starts with $0\leq x,y\leq 2$,
increments both variables by two in a loop, and requires $x\neq 4$ if
$y=0$ after the loop. In the existing run, \tool generated the invariant
that $x-y$ equals its entry value, together with a loop frame allowing
changes only to $x$ and $y$. WP discharged the invariant-establishment,
invariant-preservation, loop-frame, and original-assertion obligations.
This checks whether the generated annotations support the supplied target;
it is not a claim that all runtime-safety properties were checked.
}

\paragraph{Java/JML (SpecGen)}
We compare \tool with the SpecGen pipeline on SpecGenBench
(120 cases) and the SV-COMP Java subset (265 cases). Both systems use
Claude Sonnet~4.6 and GPT-3.5, with ten independent attempts per case
and model. OpenJML checks the generated JML method contracts and loop
annotations against the implementation.
 A case is solved when at least one attempt discharges all OpenJML obligations;
the per-attempt success rate is the fraction of attempts that do so.

On SpecGenBench, SpecGen solves more cases, whereas \tool has the higher
per-attempt success rate. With Claude, \tool solves 112 cases (88.08\%
of attempts pass) and SpecGen solves 116 (84.92\%). With GPT-3.5, the
corresponding results are 103 cases (80.00\%) and 109 (69.50\%). On the
SV-COMP Java subset, the per-attempt success rates for \tool and SpecGen
are 81.96\% and 78.04\% with Claude and 69.40\% and 50.91\% with
GPT-3.5.

\subsection{RQ4: Findings on Real-World Software}
\label{sec:rq3}

Applied to mature and heavily-fuzzed real-world software, \tool surfaced
defects in Linux drivers and eleven other open-source software projects. 

\begin{table}[t]
\centering
\small
\setlength{\tabcolsep}{3.5pt}
\renewcommand{\arraystretch}{1.08}
\caption{Fourteen defects found by \tool and fixed in Linux mainline as of
23~November~2026.}
\label{tab:rq4-linux}
\begin{tabularx}{\columnwidth}{@{}l>{\raggedright\arraybackslash}Xl@{}}
\toprule
\textbf{Driver} & \textbf{Bug description} & \textbf{Mainline commit} \\
\midrule
Amphion VPU & Extra runtime PM put on probe failure & \href{https://github.com/torvalds/linux/commit/0cabf27cf1e6f7ee5f3c5893c0f1930814d949af}{\texttt{0cabf27cf1e6}} \\
\texttt{atcspi200} & Clock disabled after suspend failure & \href{https://github.com/torvalds/linux/commit/f4b14e67baffd2c879765275bda998a183594e8b}{\texttt{f4b14e67baff}} \\
\texttt{bcm-qspi} & Hardware teardown after suspend failure & \href{https://github.com/torvalds/linux/commit/44d09ef0ba690ac46d17eb05c12a47a5dd75e231}{\texttt{44d09ef0ba69}} \\
\texttt{bcm63xx} & Clock disabled after suspend failure & \href{https://github.com/torvalds/linux/commit/bbbc1ef5b35341c5122418a14f8924688a4bb954}{\texttt{bbbc1ef5b353}} \\
\texttt{bcm63xx-hsspi} & Clock teardown after suspend failure & \href{https://github.com/torvalds/linux/commit/3b5414d76a0381db77b9694537254af7acfa187a}{\texttt{3b5414d76a03}} \\
\texttt{bcmbca-hsspi} & Clock teardown after suspend failure & \href{https://github.com/torvalds/linux/commit/faa878d4805ca375738294bc52eae310eb287b16}{\texttt{faa878d4805c}} \\
\texttt{exynos5422-dmc} & Unchecked clock enables and missing rollback & \href{https://github.com/torvalds/linux/commit/cce46902ea9fcd585cd200fffa3166f63bc571f1}{\texttt{cce46902ea9f}} \\
Exynos FIMD & Bus clock leak on resume failure & \href{https://github.com/torvalds/linux/commit/d04392ce52d33bce4d532bd8b965151ed4ce8670}{\texttt{d04392ce52d3}} \\
\texttt{fsl-dspi} & Incomplete suspend/resume cleanup & \href{https://github.com/torvalds/linux/commit/2543355f3da56a297e3803ccb5d29f4dce5f18f1}{\texttt{2543355f3da5}} \\
\texttt{gpio-omap} & Ignored debounce clock preparation failure & \href{https://github.com/torvalds/linux/commit/9df982f1f283d2f2a9aad764dff92e36ed764e16}{\texttt{9df982f1f283}} \\
\texttt{hix5hd2\_gmac} & Redundant NAPI deletion during teardown & \href{https://github.com/torvalds/linux/commit/f307a7dc32097c11413178fca437a10d20890bc2}{\texttt{f307a7dc3209}} \\
\texttt{jh7110-cryp} & Unchecked clock enable failures & \href{https://github.com/torvalds/linux/commit/5358248961ae23f7253d010ddfde1cfd2556ce22}{\texttt{5358248961ae}} \\
\texttt{nxp-fspi} & Runtime PM left enabled on probe failure & \href{https://github.com/torvalds/linux/commit/5d14285d60ba95561609a293064654a0c7d7d928}{\texttt{5d14285d60ba}} \\
\texttt{rockchip-sfc} & Runtime PM left enabled after removal & \href{https://github.com/torvalds/linux/commit/34e32e9ade58e32daf933974c9b697d066fc5aaa}{\texttt{34e32e9ade58}} \\
\bottomrule
\end{tabularx}
\end{table}

\begin{table}[t]
\centering
\caption{Sixty-eight defects found by \tool and confirmed by manual source
review. Six of the eleven projects are continuously fuzzed through
OSS-Fuzz, and account for 51 of the defects.}
\label{tab:rq4}
\normalsize
\begin{tabular}{@{}llr@{}}
\toprule
Project & Domain & Count \\
\midrule
\multicolumn{3}{@{}l}{\emph{OSS-Fuzz-integrated projects}}\\
u-boot      & bootloader: filesystem/EFI parsers & 32 \\
gpac        & multimedia framework                 & 10 \\
libbej      & Redfish BEJ decoder                  & 4 \\
jq          & JSON processor (fixed in 1.8.2)      & 2 \\
libredwg    & DWG/CAD reader                        & 2 \\
libarchive  & archive formats                      & 1 \\
\midrule
\multicolumn{3}{@{}l}{\emph{Other open-source (not in OSS-Fuzz)}}\\
libmikmod   & module-audio loaders                 & 4 \\
libmodplug  & module-audio loaders                 & 2 \\
adplug      & AdLib-audio loaders                  & 3 \\
CCC         & AI-generated Rust C compiler         & 7 \\
AWS Neuron  & Trainium/Inferentia ioctl handler    & 1 \\
\midrule
\textbf{11 projects} & & \textbf{68} \\
\bottomrule
\end{tabular}
\end{table}

For Linux,  we
submitted \textbf{42 distinct fix proposals} based on \tool's findings, of which \textbf{14
have been merged into mainline} (Table~\ref{tab:rq4-linux}). 
We reviewed each report against the kernel
source, wrote the patch, and revised it across review rounds, counting a proposal once
across its revisions.  The maintainers reviewed and accepted these patches
independently of \tool and of our own audit.  The fixes form four error-path
patterns: six fixes in SPI controller drivers correct suspend or resume
paths that previously tore down clocks or other resources after controller
suspension had failed; three fixes correct runtime-PM state or
reference-count imbalances around probe or removal; four fixes handle clock
preparation or enablement failures by propagating the error or restoring
consistent cleanup; and one fix removes a redundant NAPI deletion during
network-device teardown.

Across the other eleven projects, we confirmed \textbf{68 defects} through manual source
review (Table~\ref{tab:rq4}).
For C/C++ targets, evidence generally includes standalone sanitizer
reproducers. For CCC, a C compiler written in Rust, we use Kani counterexamples and
source review. The jq maintainers published a security
advisory for each of the two defects we reported and fixed both in
jq~1.8.2.\footnote{See
the jq security advisories
\href{https://github.com/jqlang/jq/security/advisories/GHSA-ggc9-rpv2-xgpm}
     {\texttt{GHSA-ggc9-rpv2-xgpm}} and
\href{https://github.com/jqlang/jq/security/advisories/GHSA-gvwx-xj9r-3frq}
     {\texttt{GHSA-gvwx-xj9r-3frq}}.}
We report aggregate counts for findings under coordinated disclosure.

The largest cluster, 32 defects, is in u-boot, which joined OSS-Fuzz only in
2026 and whose squashfs, btrfs, ext4, and EFI image parsers remain under-fuzzed;
several defects are reachable only when the bootloader mounts attacker-supplied
removable media, for which no fuzz harness exists yet. 

\paragraph*{Verification results as positive evidence}
Beyond defects, many functions in these hardened libraries returned a
\emph{verified clean} verdict, among them OpenSSL's ASN.1 DER decoders and
libxml2's XPath pattern compiler.  On code this heavily fuzzed, that
verdict is itself a result, verification evidence that the checked properties
hold within the unwinding bound under the inferred specification.  A
surviving counterexample is a defect report instead, and an inconclusive
query is neither. An LLM-only auditor cannot provide such a verdict.

\subsection{Threats to Validity}
\label{sec:threats}

\paragraph*{Soundness of agent decisions}
An LLM agent can propose a wrong specification, configuration, or
refinement, and the proposal itself carries no signal of that.
Table~\ref{tab:agents} pairs each proposal with the check applied to it,
and no agent output is adopted unchecked.

\paragraph*{Over-approximation} 
Postcondition stubs can admit behaviours that real callees cannot
produce. \tool checks witnesses against the implementation using callee
feasibility, reachability, dynamic replay, and a realism audit. Replay evidence comes from the analysis host rather than the target hardware. 

\paragraph*{Evaluation}
Lacking ground truth for LLM-generated code, true positives are judged by a
hybrid criterion that combines a generated reproducer, compiled and run to
trigger the defect where one can be produced, with manual source review of every
finding; the SV-COMP tasks are instead scored directly against the
given labels.

\paragraph*{External validity}
Our subjects span three deliberately different populations, namely LLM-generated
systems code, mature OSS-Fuzz-hardened libraries, and curated verification
benchmarks, chosen to probe generality. Even so, the absolute defect rates and catch
counts depend on corpus selection and may not transfer to other domains, languages,
or systems.

\section{Related Work}
\label{sec:related}

\subsection{LLM-based specification and invariant synthesis}

The premise that verification is gated on specifications is itself
long-studied~\cite{neider2024formal}.
Inferring program specifications predates LLMs: Daikon mines likely
invariants from executions~\cite{ernst2007daikon}, Houdini filters candidate
annotations for ESC/Java~\cite{flanagan2001houdini}, and others mine
specifications from how code is \emph{used}, including API-usage patterns
from execution traces~\cite{pradel2009automatic}, preconditions that hold at
a procedure's call sites~\cite{ramanathan2007static}, and temporal
specifications from object usage~\cite{wasylkowski2011mining}.

Recent work revisits this goal with LLMs.  AutoSpec~\cite{autospec} and
SpecGen~\cite{ma2025specgen} synthesize ACSL/JML contracts for deductive
verifiers in a generate--check--refine loop; Faria et
al.~\cite{faria2026automatic} add test oracles and verification annotations,
and separation-logic specifications have also been
synthesized~\cite{suresh2026agentic}.  SpecSyn~\cite{ma2026specsyn} and
C2S~\cite{zhai2020c2s} infer formal specifications from code and
natural-language comments; further studies assess how reliably LLMs produce
and reason about such specifications~\cite{le2025can,bouras2025hoareprompt}. 

A parallel line of work targets a single artifact.
Loopy~\cite{kamath2024leveraging} proposes and repairs loop invariants
against a checker, Wu et al.~\cite{wu2024llm} and Pirzada et
al.~\cite{pirzada2024llm} pair LLM-generated invariants directly with
bounded model checking, and AutoVerus~\cite{yang2025autoverus} generates the
proof annotations Verus needs for Rust, while Safe4U~\cite{li2025safe4u}
flags unsound safe encapsulations of unsafe code.

These systems produce specifications, invariants, or proofs for a downstream
tool; \tool instead treats specification inference as one stage of a
verification workflow, integrating caller-grounded contracts with bounded
model checking, counterexample validation, and compositional refinement.

\subsection{Agentic verification and LLM bug finding}

Several systems pair LLM agents with formal reasoning. \fmagent~\cite{fmagent}
infers Hoare-style contracts and checks them with LLM reasoning;
Lemur~\cite{lemur} has LLMs synthesize artifacts checked by external engines;
ConVer~\cite{pirzada2026conver} scales deductive verification from LLM-generated
contracts given a user-provided property; and Code-Augur~\cite{luo2026codeaugur}
and specification-guided detection~\cite{DBLP:journals/pacmse/ZhuLGQDLWWHL26} expose security
assumptions as assertions and flag violating code. \fmagent and Code-Augur are
closest to \tool (all infer and refine specifications) but differ in the
\emph{verification oracle}. \fmagent relies on LLM reasoning, Code-Augur on 
guided fuzz testing, 
and LLM-powered symbolic execution~\cite{li2025large} on an approximate
engine, whereas \tool discharges each specification with bounded model
checking, exhaustively covering inputs up to a bound and yielding concrete
counterexamples and bounded verification verdicts.

A related line of work uses LLMs for interactive proof
construction~\cite{polu2020generative, first2023baldur, yang2023leandojo,
tu2025agentic} and for bug and vulnerability
detection~\cite{ding2024vulnerability}, which yield proofs or bug reports
rather than the model-checked verdicts \tool produces.

\subsection{Model checking, program analysis, deductive verification}

Bounded model checkers~\cite{cbmc, kani, cordeiro2018jbmc, cordeiro2011smt}
verify whether a program satisfies some specification up to a given depth,
and tools such as SeaHorn~\cite{gurfinkel2015seahorn},
CPAchecker~\cite{beyer2011cpachecker}, and BLAST~\cite{blast} scale
verification via abstraction, Horn-clause reasoning, and CEGAR.  SLAM/SDV
checks fixed driver-API properties~\cite{slam}, Smatch~\cite{smatch}
performs pattern-based kernel analysis, and Infer~\cite{calcagno2011infer}
scales compositional bug finding through procedure summaries.

Deductive verifiers such as Boogie~\cite{barnett2005boogie},
Dafny~\cite{leino2010dafny}, Why3~\cite{filliatre2013why3},
Frama-C/WP~\cite{framac}, and Verus~\cite{lattuada2023verus} give strong
guarantees but require substantial user-supplied annotations.  These systems
excel once properties exist; \tool targets the complementary problem of
inferring and refining the specifications that make them applicable to
unannotated code, while retaining the concrete witnesses produced by bounded
model checking.

\subsection{Symbolic execution and fuzzing}

KLEE~\cite{klee} and EXE~\cite{cadar2008exe}, concolic engines
(DART~\cite{godefroid2005dart}, CUTE~\cite{sen2005cute}), and hybrid
symbolic-fuzzing techniques~\cite{godefroid2008automated, cha2012unleashing,
poeplau2020symbolic} expose bugs through path exploration. Large-scale
fuzzing infrastructure such as OSS-Fuzz~\cite{chang2016oss} and
AFL++~\cite{fioraldi2020afl++} has hardened many libraries we evaluate.
  Serval~\cite{serval} symbolically
evaluates systems software within decidable fragments.
Unlike \tool, these generate inputs rather than per-function specifications
or bounded correctness verdicts, which makes them complementary to \tool.

\subsection{Compositional verification, CEGAR}

\tool builds on assume-guarantee reasoning~\cite{assume-guarantee} and
CEGAR~\cite{cegar}. Function contracts act as assume-guarantee obligations,
and unrealistic counterexamples trigger refinement at the specification
level, propagated compositionally across the call graph.

\section{Conclusion}
\label{sec:conclusion}
We have presented \emph{agentic model checking} and \tool, which place LLM
agents around a bounded model checker using the principle that \emph{agents
propose, solvers verify}: the agents take on the semantic judgement that a
checker lacks, while a deterministic formal checker establishes results
for the resulting bounded queries. The central idea is to let an agent \emph{orchestrate} the
work of the checker rather than fix a single recipe, driving the formal
engine to different operating points so it scales to large, multi-function
code that a monolithic approach cannot solve.  Across LLM-generated systems
code, verification benchmarks, and hardened libraries, \tool finds confirmed
defects in heavily tested code, including bugs in real drivers that are
beyond the reach of monolithic model checking.  It also produces bounded
verification results.  More broadly, our experiments suggest that LLM agents
are most useful in verification not as a replacement for the formal engine
but as the \emph{orchestrator} that decides how to apply it.
%

\section*{Data Availability}
The BMC-Agent source code, benchmark materials, evaluation data, and
reproduction instructions are publicly available at
\url{https://github.com/agentic-prover/aprover/tree/paper-artifacts}.

\section*{Acknowledgement}
\label{sec:acknowledgement}

Research reported in this paper was partially supported by an Amazon
Research Award, Fall 2025.  Any opinions, findings, and conclusions or
recommendations expressed in this material are those of the authors and do
not reflect the views of Amazon.

\commentout{
\smallskip\noindent\textbf{Future work.}
The most important next steps are
(i) a broad quantitative evaluation establishing precision and recall at
scale, with controlled ablations of each pipeline stage and head-to-head
comparison against BMC-alone baselines;
(ii) a deeper answer to the spec-correctness assumption, through spec
triangulation across multiple sources and a measured audit of LLM-inferred
references;
(iii) evaluating \tool on historical buggy corpora to assess
generalisation beyond LLM-generated code; and
(iv) reducing token cost via spec-store reuse and stage-level distillation,
so that whole-codebase sweeps amortise across runs rather than paying full
cost each time.
}

\raggedbottom
\IEEEtriggeratref{45}
\IEEEtriggercmd{\balance}
\bibliographystyle{IEEEtran}
\bibliography{references}

\end{document}